\documentclass[letterpaper]{article} 
\usepackage{aaai2026}  
\usepackage{times}  
\usepackage{helvet}  
\usepackage{courier}  
\usepackage[hyphens]{url}  
\usepackage{graphicx} 
\usepackage{natbib}  
\usepackage{caption} 
\usepackage{algorithm}
\usepackage{algorithmic}

\usepackage{booktabs}
\usepackage{amsmath}
\usepackage{amsthm}
\usepackage{amssymb}
\usepackage{multirow}
\usepackage{xcolor}
\usepackage{bm}
\usepackage{subfig}
\usepackage{diagbox}
\usepackage[switch]{lineno}
\usepackage[table]{xcolor}
\newcolumntype{g}{>{\columncolor{gray!30}}c}

\usepackage{newfloat}
\usepackage{listings}
\DeclareCaptionStyle{ruled}{labelfont=normalfont,labelsep=colon,strut=off} 
\floatstyle{ruled}
\newfloat{listing}{tb}{lst}{}
\floatname{listing}{Listing}
\title{Reference Recommendation based Membership Inference Attack against Hybrid-based Recommender Systems}
\author{
    Xiaoxiao Chi\textsuperscript{\rm 1}, Xuyun Zhang\textsuperscript{\rm 1}\footnote{Corresponding Author. Contact xuyun.zhang@mq.edu.au or Hongsheng.Hu@newcastle.edu.au}, Yan Wang\textsuperscript{\rm 1}, Hongsheng Hu\textsuperscript{\rm 2*}, Wanchun Dou\textsuperscript{\rm 3}
}
\affiliations{
    \textsuperscript{\rm 1}Macquarie University
    
    \textsuperscript{\rm 2}The University of Newcastle 
    
    \textsuperscript{\rm 3}Nanjing University 

}

\usepackage{bibentry}

\begin{document}

\maketitle

\begin{abstract}
Recommender systems have been widely deployed across various domains such as e-commerce and social media, and intelligently suggest items like products and potential friends to users based on their preferences and interaction history, which are often privacy-sensitive. Recent studies have revealed that recommender systems are prone to membership inference attacks (MIAs), where an attacker aims to infer whether or not a user's data has been used for training a target recommender system. However, existing MIAs fail to exploit the unique characteristic of recommender systems, and therefore are only applicable to mixed recommender systems consisting of two recommendation algorithms. This leaves a gap in investigating MIAs against hybrid-based recommender systems where the same algorithm utilizing user-item historical interactions and attributes of users and items serves and produces personalised recommendations. To investigate how the personalisation in hybrid-based recommender systems influences MIA, we propose a novel metric-based MIA. Specifically, we leverage the characteristic of personalisation to obtain reference recommendation for any target users. Then, a relative membership metric is proposed to exploit a target user's historical interactions, target recommendation, and reference recommendation to infer the membership of the target user's data. Finally, we theoretically and empirically demonstrate the efficacy of the proposed metric-based MIA on hybrid-based recommender systems.
\end{abstract}


\section{Introduction}

Recommender systems aim to provide users with items that match their interests, which have become indispensable part in various domains~\cite{zhang2019deep}, such as e-commerce~\cite{ugurlu2025style4rec}, entertainment~\cite{stiennon2020learning}, and social media~\cite{wei2019mmgcn,gao2023survey}. The success of recommender systems can be contributed to two main factors: the rapid advancement of recommendation algorithms and the availability of large-scale data of users and items. Recommender systems can learn users' preference towards items basically from two kinds of data, i.e., the user-item interactions such as ratings or buying behaviors~\cite{sun2019bert4rec}, and the attribute information about the users and the items such as textual profiles or relevant keywords~\cite{volkovs2017dropoutnet}.

Despite their effectiveness, recommender systems have been shown to be vulnerable to membership inference attacks (MIAs)~\cite{zhang2021membership,wang2022debiasing,chi2024shadow}, which pose significant privacy risks. In such attacks, an adversary aims to determine whether a particular user's data was used to train a target recommender system. This is achieved by exploiting behavioral differences in the system's responses to users whose historical data was included in training versus those whose data was not. Such distinctions can lead to the leakage of membership information, thereby violating privacy regulations such as the GDPR~\cite{rosen2011right} and CCPA~\cite{pardau2018california}, both of which emphasize the protection of individual data privacy.

While prior works have explored membership inference attacks (MIAs) on recommender systems, they often rely on an impractical assumption: all users with historical interactions are treated as members, whereas non-members are assumed to be entirely new users with no interactions. This assumption does not hold in real-world scenarios, where an existing user may be a non-member if their data was excluded from the training process. For example, a user might have opted out of data collection due to privacy preferences or joined the platform before the data collection window used for model training. Moreover, existing MIAs primarily focus on recommendation settings involving mixed components, where the system exhibits distinct behaviors for existing users versus new users without interaction histories. As a result, there remains a critical gap in understanding MIAs in hybrid-based recommender systems, where a unified algorithm serves both existing and new users.

In this paper, we investigate the privacy vulnerabilities of hybrid-based recommender systems through the lens of MIAs. Hybrid recommenders utilise both users' historical interactions and attribute information, enabling them to generate personalised recommendations even for users with no interaction history. This challenges the assumptions of existing MIA methods, which rely on behavioral differences between users with and without historical data. To address this, we propose a novel reference recommendation-based MIA that exploits the personalisation characteristics of hybrid-based recommenders. Specifically, we introduce an additional query to the target recommender system that retrieves a reference recommendation for a target user based solely on their attributes. We then define a relative membership metric that leverages the target user's historical interactions, the actual recommendation, and the reference recommendation to infer membership status. If the value of this metric falls below a predefined threshold, the user is classified as a member; otherwise, as a non-member. Beyond attack design, we evaluate the privacy risk of hybrid recommender systems when differential privacy (DP) is applied. Experimental results show that DP provides a level of privacy protection, while our attack remains consistently effective. Our contributions are summarized as follows:

\begin{itemize}
     \item We present the \textit{first} study of membership inference attacks (MIAs) on hybrid-based recommender systems. We propose a novel reference recommendation based MIA that leverages the personalisation characteristics of hybrid-based recommenders, and introduce a relative membership metric to effectively and efficiently infer membership status.
     
    \item We provide a detailed mathematical analysis of the metric and demonstrate that its nonlinear formulation enables more effective discrimination in membership compared to SOTA methods.
    
    \item We conduct comprehensive experiments on benchmark datasets and recommender systems. The results show that our proposed attack consistently outperforms baseline methods in both effectiveness and efficiency.

\end{itemize}

\section{Related Work}
\noindent \textbf{Membership Inference Attacks.} Membership inference attacks~\cite{hu2022membership,carlini2021extracting,sablayrolles2019white} have become standard privacy attacks to evaluate privacy leakage of machine learning models. In the context of recommender systems, \citet{zhang2021membership} investigate the first MIAs on such models. Specifically, they leverage the well-established shadow training techniques~\cite{shokri2017membership,long2020pragmatic} to train a shadow recommender system to mimic the behavior of the target recommender system. Then, representative meta-data about the training samples (i.e., members) and non-training samples (i.e., non-members) is collected from the trained shadow model, which will be used to train a binary attack classifier. Following this attack pipeline, other works~\cite{wang2022debiasing,zhu2023membership,zhong2024interaction} further investigate MIAs on different types of recommender systems, e.g., sequential recommender systems, and improve the attack performance by mitigating the discrepancy between the shadow model and the target model. 

\noindent \textbf{Hybrid-based Recommender Systems.}
Recommender systems~\cite{resnick1997recommender} analyse users' data and preferences to provide recommendations. Among various recommendation algorithms~\cite{zhang2019deep}, traditional collaborative filtering (CF) algorithms~\cite{koren2021advances} such as Neural Collaborative Filtering~\cite{he2017neural} serve as representative methods, making recommendations through learning from user historical interactions. However, these CF algorithms suffer from user cold-start problems~\cite{zhang2025cold,zhu2025addressing,lan2025next}, where they cannot make personalised recommendations to new users who do not have any historical interactions with the model. To address user cold-start problems, hybrid-based recommender systems~\cite{volkovs2017dropoutnet,zhu2020recommendation} focus on exploiting attribute information to alleviate the problem. Specifically, the historical interactions and attributes of the users and items are utilized to train the hybrid-based recommender system. After training the model, the hybrid-based recommender system can make personalised recommendations to new users {based on their attribute information}.

\section{Preliminaries and Problem Formulation}

\noindent \textbf{Preliminaries.}
Traditional collaborative filtering can recommend items utilizing user historical interactions, e.g., shopping transactions and movie ratings~\cite{malitesta2025formalizing}, but suffers from the user cold-start problem and fails to make effective recommendations for new users without any interactions. Instead, hybrid-based approaches such as DropoutNet~\cite{volkovs2017dropoutnet} exploit both the interactions and the attribute information of users and items, e.g., user age and item description, which are often available in real applications of recommender system. Such information can empower the hybrid recommenders to achieve a more informed recommendation and address the cold-start problem. Our work mainly concentrates on this common and realistic setting in recommender systems.

Let $\mathcal{U}$ and $\mathcal{I}$ denote the set of users and items, respectively, and a matrix $\mathcal{R}^{|\mathcal{U}|\times|\mathcal{I}|}$ describe the users' historical interactions with items. Each entry of $\mathcal{R}$ is a domain-specific encoding value for an interaction, e.g., a binary value representing if a user has bought an item or not in a shopping transaction. For a user $u~{\in}~\mathcal{U}$, its attributes can be represented as a vector $\Phi_{u}$, with domain-specific encoding values, e.g., a gender attribute value can be encoded as an integer. Furthermore, we use a matrix $\Phi^{\mathcal{U}}$ to denote the attribute information for all users, with each row for one user. Similarly, we can have the vector $\Phi_{i}$ for the attribute information of an item $i~{\in}~\mathcal{I}$ and the matrix $\Phi^{\mathcal{I}}$ for all the items.

The hybrid-based recommender system, denoted as $\mathcal{RS}$, exploits $\Phi^{\mathcal{U}}$, $\Phi^{\mathcal{I}}$, and $\mathcal{R}^{|\mathcal{U}|\times|\mathcal{I}|}$ to train a recommendation model. It takes both attributes and interactions as the input. Since a user's interactions $\mathcal{R}_u$ are often of high dimensionality and extreme sparsity in real applications, a latent vector which is derived from the interactions as a dense representation $\mathcal{P}_u$, is usually employed to feed the hybrid-based recommenders instead, e.g., DropoutNet~\cite{volkovs2017dropoutnet} and Heater~\cite{zhu2020recommendation}. When making predictions for a user $u$, the recommender $\mathcal{RS}$ receives the user information $\langle \mathcal{P}_u , \Phi_u   \rangle $ as input, and outputs a list of items $\mathcal{Y}_u$ as recommendation. In particular, when $u$ is a new user, who has no interactions with items at all, the hybrid-based recommender can still conduct personalised recommendation with the user attribute information only, while traditional collaborative filtering methods fail to handle new users effectively and suffer from the cold-start problem.

\noindent \textbf{Problem Formulation.}
Like other machine learning models, a recommender system is also prone to membership inference attacks (MIAs), as demonstrated in the recent works~\cite{zhang2021membership,wang2022debiasing, chi2024shadow}. In such an attack, given a user $u$'s historical interactions $\mathcal{R}_u$ and attribute information $\Phi_u$, an attacker aims to determine whether or not this user's data has been used for training a recommender system $\mathcal{RS}$, which is referred to as the target model. If yes, the user's data is referred to as a member; otherwise, as a non-member.
Our work focuses on the black-box attack setting where the attacker can only obtain recommendations from $\mathcal{RS}$, which is more practical yet more challenging in real-world applications because machine learning models are often deployed as services without exposing internal implementation \cite{7424435}. With certain prior knowledge $\mathcal{K}$, an MIA method $\mathcal{A}$ can be formally described as follows: 
\begin{equation}
    \mathcal{A}: \mathcal{R}_u, \Phi_u, \mathcal{RS}, \mathcal{K} \rightarrow \{\textit{member}, \textit{non-member}\}.
\end{equation}

\noindent \textbf{Research Gaps.} The works~\cite{zhang2021membership,wang2022debiasing} initially investigated MIAs against recommender systems, where the recommender systems they targeted are in essence a mix of two recommendation methods: collaborative filtering for members and popularity-based recommendation for non-members. They have an implicit assumption that all existing users with interactions are members, and non-members are new users without any interactions. However, this assumption is impractical in real applications because an existing user's data can be a non-member if it was not used for training a recommender system. Also, this assumption implies that the recommender is aware of the membership of a user.

Instead, it is more common that a recommender system of mixed recommendation components can behave differently between existing users (either members or non-members) and new users (certainly non-members) as explored in the work~\cite{chi2024shadow}, where collaborative filtering is for existing users and popularity-based recommendation is for new users. A highly efficient shadow-free attack method is proposed in this work to determine membership directly without training a shadow model. Besides comparing the recommendation results with the historical interactions, the comparison with the most popular items from popularity-based recommendation is also used for membership inference and improves the attack performance. However, as shown above, existing MIAs on recommenders are limited to the scenarios involving mixed recommendation algorithms, leaving a gap in addressing hybrid-based systems where the same algorithm serves both existing users and new users.

Since both historical interactions and user attributes are used, hybrid-based recommender systems produce more personalised recommendation, even for new users without any historical interactions. An interesting question naturally arises: \textit{How does personalisation in hybrid-based recommender systems influence MIAs?} This is a non-trivial question. On the one hand, more personalisation may imply more privacy exposure, aggravating membership privacy disclosure. On the other hand, hybrid-based recommender systems can address the cold-start problem well and mitigate the overfitting issue, which should strengthen the defence to MIAs as observed on other machine learning models~\cite{li2021membership}. To answer this question, a follow-up task is to propose an effective and efficient  MIA on hybrid-based recommender systems. It is worth noting that the efficient shadow-free method developed in~\cite{chi2024shadow} cannot be applied herein because new users receive different recommendations specific to their own attributes. Thus, designing a novel MIA method to explore how personalisation affects membership inference motivates our research herein.

\begin{figure*}[t]
\centering
\includegraphics[width=0.9\linewidth]{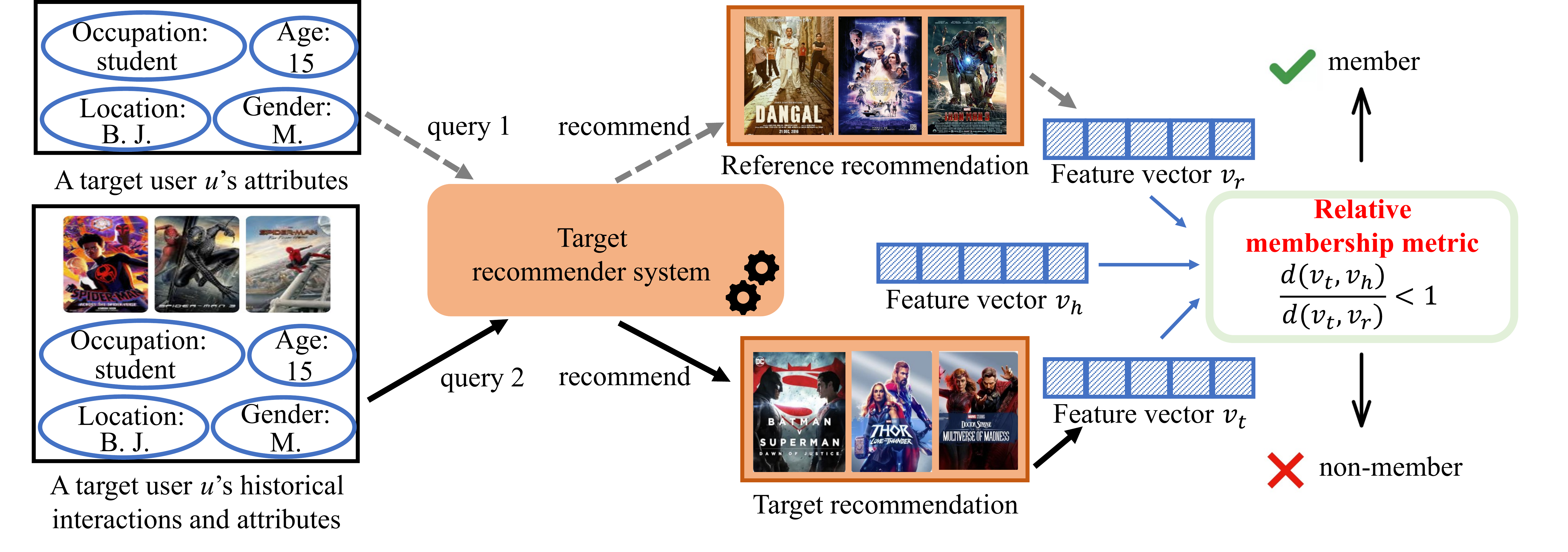}
\caption{ An example showing the workflow of our proposed reference recommendation based MIA. }
\label{fig:mia}
\end{figure*}

\noindent \textbf{Threat Model.} 
Our work assumes that an attacker does not know the data distribution of target dataset and the model architecture of target recommender system $\mathcal{RS}$. Following the convention in previous works~\cite{zhang2021membership,wang2022debiasing,chi2024shadow}, we assume the attacker can collect and obtain a dataset of user-item historical interactions where elements in item set $\mathcal{I}$ are all included. This dataset can be obtained by crawling the Internet, from platforms such as Douban~\cite{zhu2023membership}. With the dataset, the attacker can use matrix factorization~\cite{koren2009matrix} to obtain item feature embeddings that will be utilized to calculate distance in membership inference.

\section{Methodology}

\subsection{Reference Recommendation based MIA: Overview}

The intuition behind prior attacks on recommender systems~\cite{zhang2021membership,wang2022debiasing} is to exploit the distance between the feature vectors of the target user's recommendation and historical interactions, denoted as $\bm{v}_t$ and $\bm{v}_h$ respectively. Then, the difference, denoted as $d(\bm{v}_t, \bm{v}_h)$, has been used as the key information for membership measurement. As analysed in the research gaps, its effectiveness in existing works mainly stems from the employment of two recommendation algorithms for members and non-members. In a hybrid-based recommender system, however, a single algorithm is used to make personalised recommendation for all users. As a result, the traditional membership metric $d(\bm{v}_t, \bm{v}_h)$ will become less distinguishable between members and non-members, which can invalidate the effectiveness of existing attacks. 

To address this challenge, we propose a simple yet effective method to amplify the difference between members and non-members based on a recommender system's specific property. The core idea is to perform an extra query only with the target user's attribute information to produce a reference recommendation, and the difference between the target recommendation and the reference recommendation, denoted as $d(\bm{v}_t, \bm{v}_r)$, is used to rescale $d(\bm{v}_t, \bm{v}_h)$. This leads to a novel membership measurement for hybrid-based recommender systems, named as the relative membership metric. Formally, for a target user $u$, it is denoted as $\rho(u)$ and defined as follows:
\begin{equation}
    \rho(u) = \frac{d(\bm{v}_t, \bm{v}_h)}{d(\bm{v}_t, \bm{v}_r)}.
    \label{eq:metric}
\end{equation}

After obtaining the metric value $\rho(u)$, we can determine the membership status of $u$ by comparing $\rho(u)$ and $1$. The user $u$ is predicted as a member if $\rho(u) < 1$, or a non-member otherwise. The rationale is that if the target recommendation is closer to the historical interactions than the reference recommendation, it is more likely that the historical interactions have been used in model training, and vice versa. This reference recommendation based membership inference attack is illustrated by an example in Figure~\ref{fig:mia}, where an attacker makes black-box queries to the target model, and exploits feature vectors to infer the membership status of the target user $u$'s data through the proposed relative membership metric.

\subsection{Advantages of Relative Membership Metric}

As the core of the reference recommendation based MIA, the proposed relative membership metric has demonstrated several salient features. With a simple comparison with $1$, our metric is user-friendly and its use does not require a user-specified and domain-specific threshold to make the membership decision, which is necessary in many existing metrics. More importantly, as our metric is a rescaled value, it is less sensitive to the absolute values of $d(\bm{v}_t, \bm{v}_h)$ that is often exploited in existing metrics. Instead, it is determined whether target recommendation is more similar to the historical interactions or the reference recommendation, which is the essential information for membership decision. In this regard, the challenge brought by the personalisation in hybrid-based recommender systems can be well addressed by this relative metric. 

\begin{figure}[t!]
    \centering
  \subfloat[Metric Functions]{%
       \includegraphics[width=0.23\textwidth]{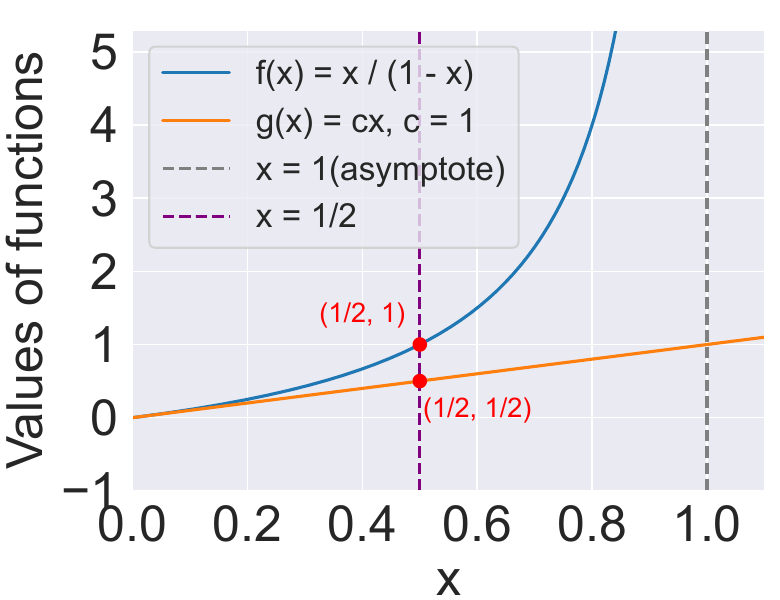}}
  \subfloat[Metric Derivative Functions]{%
        \includegraphics[width=0.23\textwidth]{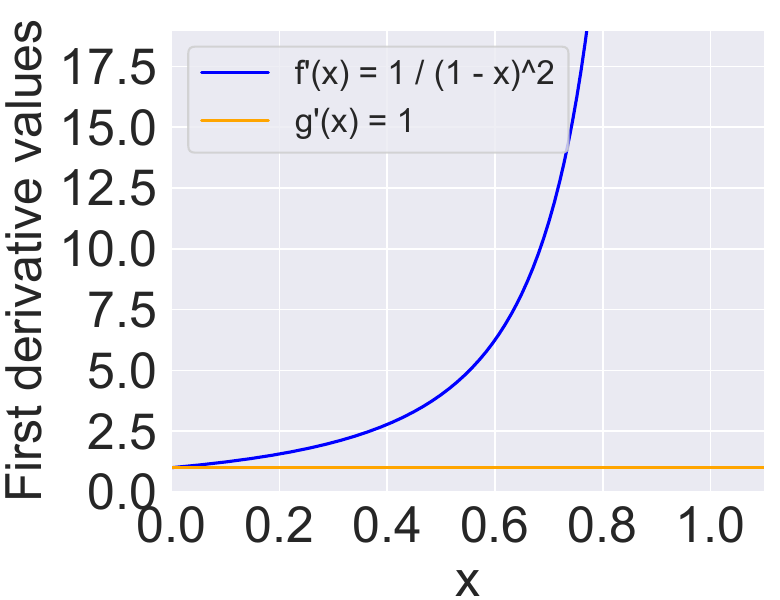}}
  \caption{Comparison of the metric functions $f(x)$ and $g(x)$.}
  \label{fig:math}
\end{figure}

The non-linear functional form of our relative metric empowers it to be more distinguishable on membership than the existing metrics which are usually linear to $d(\bm{v}_t, \bm{v}_h)$. If a user $u$ with historical interactions $\bm{v}_h$ is a member, the target recommendation vector $\bm{v}_t$ is closer to $\bm{v}_h$. An extreme case is $d(\bm{v}_t, \bm{v}_h)=0$ when the recommender returns a result the same as the historical interactions, which is possible in a heavily overfitted model. If $u$ is a non-member, the target recommendation vector $\bm{v}_t$ is closer to the reference recommendation $\bm{v}_r$. The value $d(\bm{v}_t, \bm{v}_h)$ increases when $\bm{v}_t$ approaches $\bm{v}_r$, with its maximum value $\lim_{\bm{v}_t\to\bm{v}_r}d(\bm{v}_t, \bm{v}_h)=M$, where $M$ is the upper bound of $d(\bm{v}_t, \bm{v}_h)$. Then, we can estimate $d(\bm{v}_t, \bm{v}_r)$ by $M-d(\bm{v}_t, \bm{v}_h)$ because $\bm{v}_t$ should be either closer to $\bm{v}_h$ or $\bm{v}_r$. Let $x=\frac{d(\bm{v}_t, \bm{v}_h)}{M}$ be a normalised variable, our relative membership metric can be equivalent to the function $f(x)=\frac{x}{1-x}$, where $x\in[0,1)$. Then, the range of the metric is $[0, +\infty)$. The threshold splitting membership is $f(x)=1$, and the corresponding boundary is $x=\frac{1}{2}$, implying that $\bm{v}_t$ is equally distant to $\bm{v}_h$ and $\bm{v}_r$. Our metric ranges much wider than existing metrics which are often linear to $d(\bm{v}_t, \bm{v}_h)$, which can be equivalent to a linear function $g(x)=cx$ when letting the variable $x\in[0, M)$ denote the distance $d(\bm{v}_t, \bm{v}_h)$, where $c$ is a constant. By analysing the perperties of $f(x)$ below, we further demonstrate the advantages of metric in a formal way.

The membership value changes non-linearly across members and non-members in our metric, strengthening the metric distinguishability. Since the derivative $g'(x)=c$, the changes in existing metrics stay constant across the boundary, within members, and within non-members. On the contrary, because $f'(x)=\frac{1}{(1-x)^2}>0$ and $f''(x) = \frac{2}{(1-x)^3}>0$, the metric value changes are significantly different for membership and non-members, and the changes are increasingly drastic when the target recommendation approaches the reference one. The enlarged gap between members and non-members is leveraged in our relative metric to distinguish them more effectively. The metric equivalent functions and their derivative functions are visualised and compared in Figure~\ref{fig:math} to illustrate the non-linear changes of our metric.

By revisiting the efficient shadow-free method proposed in ~\cite{chi2024shadow}, we found that it can be regarded as an application of our relative membership metric in a special case where the reference recommendation $\bm{v}_r$ is a constant value determined by the item popularity and used for all users. It is worth noting that $\bm{v}_r$ in our metric is from personalised recommendation with the use of user attributes.

\subsection{Reference Recommendation based MIA}
Before launching the reference recommendation based MIA, an attacker can use the crawled dataset to obtain the feature embedding vectors of each item to construct the feature vectors for historical interactions, target recommendation, and reference recommendation. Let $\bm{C}^{p\times q}$ be the user-item matrix which contains $p$ users' ratings of $q$ items in the crawled dataset. Using the matrix factorization technique ~\cite{koren2009matrix}, the matrix $\bm{C}^{p \times q}$ can be approximated by the product of two lower dimensional matrices denoted as $\bm{H}^{p \times l}$ and $\bm{W}^{q \times l}$, where $l$ is the number of features in the latent space. The approximate matrix denoted as $\hat{\bm{C}}^{p \times q}$ can be calculated by: 

\begin{equation}
    \hat{\bm{C}}^{p \times q} = \bm{H} \cdot \bm{W}^{T}.
\end{equation}

 The approximation can be optimised by minimising the loss function $ L(\bm{H,W}) =   ||{\bm{C}^{p \times q}} - \hat{\bm{C}}^{p \times q} ||_2$, where $||\cdot||_2$ is the Euclidean distance. After such approximate decomposition, each row of $\bm{H}$ is the latent feature of a user, and each row of $\bm{W}$ is the latent feature of an item. Let the row vector $\bm{w}_i$ be the latent feature of the item $i$. Then, $\bm{W}$ can be further denoted as $\bm{W} = 
(\bm{w}_1; \bm{w}_2; ...; \bm{w}_q)$. Based on such information, the MIA is detailed in the subsequent three steps.

\paragraph{Step i) Obtaining Target Recommendation.}
For a target user $u$, the attacker queries the target hybrid-based recommender system $\mathcal{RS}$ with $u$'s historical interactions $\mathcal{R}_u = [h_1, \cdots, h_m]$ and attributes $\Phi_u$. Then, $\mathcal{RS}$ generates $n$ recommendations $\mathcal{Y}_{u\_target}$ for $u$, denoted by:
\begin{equation}
    \mathcal{Y}_{u\_target} = [y_{t_1}, \cdots, y_{t_n}].
\end{equation}

\paragraph{Step ii) Obtaining Reference Recommendation.}
In this step, the attacker only uses $\Phi_u$ to perform a query to $\mathcal{RS}$. Being a hybrid-based model, $\mathcal{RS}$ can return personalised recommendation $\mathcal{Y}_{u\_ref}$ as the reference, denoted by
\begin{equation}
    \mathcal{Y}_{u\_ref} = [y_{r_1}, \cdots, y_{r_n}].
\end{equation}

Following the conventional setting~\cite{zhang2021membership}, the attacker can derive the feature vectors of $\mathcal{R}_u$, $\mathcal{Y}_{u\_target}$, and $\mathcal{Y}_{u\_ref}$, respectively, by taking the average of the feature vectors of the items in an interaction list as follows:

\begin{equation*}
    \bm{v}_h = \frac{1}{m} \sum_{i=1}^{m} \bm{w}_{h_i};  
    \bm{v}_t = \frac{1}{n} \sum_{i=1}^{n} \bm{w}_{y_{t_i}};
    \bm{v}_r =  \frac{1}{n} \sum_{i=1}^{n} \bm{w}_{y_{r_i}}.
\end{equation*}

\paragraph{Step iii) Inferring Membership Privacy.} To determine membership status of the target user $u$'s data, the attacker calculates the value of $\rho(u)$ according to (2), where a concrete distance measure is need. Herein we follow the prior work~\cite{chi2024shadow} to adopt the Euclidean distance for measuring the distance between two different feature vectors. It is worth noting that our relative membership metric is generic with respect to distance measurement, and other distances are applicable as well, e.g., the Jaccard distance measuring set similarity~\cite{levandowsky1971distance} and the KL divergence quantifying distribution discrepancy ~\cite{van2014renyi} can be used for the scenario where recommendation results are modeled as sets of items and item distributions, respectively. As the focus herein is on the attack methodology, we leave the investigation of other distance measures as a future work. Then, the attacker can calculate the specific metric $\rho(u)$ by 

\begin{equation}
    \rho(u) = \frac{||\bm{v}_t - \bm{v}_h||_2}{||\bm{v}_t - \bm{v}_r||_2}.
\end{equation}

By comparing $\rho(u)$ with predefined threshold $1$, the membership inference attack $\mathcal{A}(u)$ can be achieved by the following decision rule: 

\begin{equation}
\mathcal{A}(u) = \begin{cases}\textit{member}   \quad \quad {\textrm{if} \; \rho(u) < 1, }\\
\textit{non-member} \quad {\textrm{otherwise}}.
\end{cases}
\end{equation}

It is worth noting that the proposed metric allows for flexible thresholds, i.e., values other than 1, to tune sensitivity. The time complexity of our reference recommendation based MIA is $\mathcal{O}(l)$, where $l$ is the feature vector length.

\section{Experiments}

\subsection{Experimental Setup}
\noindent \textbf{Target Recommender Systems and Datasets.}
We employ two representative hybrid-based recommender systems, i.e., DropoutNet~\cite{volkovs2017dropoutnet} and Heater~\cite{zhu2020recommendation}, as target recommender systems. Benchmark datasets MovieLens-1M (ML-1M)~\cite{harper2015movielens} and MovieLens-100K (ML-100K)~\cite{harper2015movielens} are used as target datasets. For baseline methods based on shadow training, we use the ACM RecSys 2017 Challenge dataset (RecSys)~\cite{abel2017recsys} as the shadow dataset, which originates from a data distribution distinct from the target datasets. These datasets all contain user-item interactions and attributes of users and items, serving as appropriate datasets to simulate hybrid-based recommendation scenarios. Specifically, for user-item interactions, we assign a value of 1 to user-item pairs if there are interactions; otherwise, we assign 0. For ML-1M and ML-100K, each user is associated with attribute information such as age, gender, occupation, and zipcode, and each item is associated with its title and genres. For RecSys, each user is with education, location, work experience, and position, and each item has title, location, and career level. Following~\cite{volkovs2017dropoutnet}, we process and transform all categorical inputs into 1-of-n representation.

\noindent \textbf{Baseline Details.} We compare the proposed attack with two existing shadow training based attack methods: ST-MIA~\cite{zhang2021membership} and DL-MIA~\cite{wang2022debiasing}. ST-MIA is the first proposed MIA against recommender systems. Specifically, an attacker builds a shadow model to mimic the behavior of target recommender system, and constructs training data for binary attack classifier. DL-MIA follows shadow training process and improves attack performance through debiasing learning. It proposes a VAE based disentangled encoder to reduce the gap between a shadow model and a target model. To fairly compare our method with existing methods, baseline attacks are conducted under the setting where an attacker has no knowledge about the training data distribution and target model architecture.

\noindent \textbf{Evaluation Metrics.} We evaluate the effectiveness and efficiency performance of the attack. In terms of effectiveness, we adopt attack success rate (ASR) which is measured as the fraction of data that are correctly inferred, and true positive rate at low false positive rate (TPR@FPR) as evaluation metrics. Regarding efficiency, we record the overall computational time of attack methods.

\noindent \textbf{Attack Settings.} For our proposed method, we focus on four attack settings: (Dro., 1M) represents DropoutNet as the target recommender system and ML-1M as the target dataset; (Dro., 100K) represents DropoutNet as the target recommender system and ML-100K as the target dataset. (Hea., 1M) represents Heater as the target recommender system and ML-1M as the target dataset; (Hea., 100K) represents Heater as the target recommender system and ML-100K as the target dataset. For shadow training based baseline attacks, the shadow dataset is set as RecSys, and Heater as the shadow model when DropoutNet is the target model, and vice versa. Note that each experiment is repeated five times and the average results are reported. All experiments are performed on a server with NVIDIA V100 GPU and an equipment with RTX 2060. We implement attack method using Python 3.9 and PyTorch 2.2.2. We provide the implementation code in the appendix for reproducibility purposes.

\subsection{Efficacy of the Proposed Attack}

\begin{table}[]
    \centering
    
    \resizebox{0.8\linewidth}{!}{
    
    \begin{tabular}{cc|c|c|c}
    \toprule
        Target RS  & \multicolumn{2}{c|}{DropoutNet} & \multicolumn{2}{c}{Heater}  \\
        \cmidrule{2-5} 
         
        Target Dataset & ML-1M  & ML-100K & ML-1M & ML-100K \\
        \midrule 
        Our Method & 0.9340 & 0.9098 & 0.8376 & 0.7519 \\
        \cmidrule{2-5} 
        ST-MIA & 0.4995 & 0.5079 & 0.5536 &0.4920 \\
        \cmidrule{2-5} 
        DL-MIA & 0.5139 & 0.5011 & 0.4995 & 0.5 \\
        \bottomrule
    \end{tabular}
    }
    \caption{ASR of the proposed attack and two baselines.}
    \label{tab:asr}
    
\end{table}

\begin{table}[]
    \centering
    
    \resizebox{0.8\linewidth}{!}{
    
    \begin{tabular}{cc|c|c|c}
    \toprule
        Target RS  & \multicolumn{2}{c|}{DropoutNet} & \multicolumn{2}{c}{Heater}  \\
        \cmidrule{2-5} 
         
        Target Dataset & ML-1M  & ML-100K & ML-1M & ML-100K \\
        \midrule 
        Our Method & 99.84\% & 68.88\% & 97.83\% & 56.05\% \\
        \cmidrule{2-5} 
        ST-MIA & 24.61\% & 21.26\% & 25.05\% & 3.18\% \\
        \cmidrule{2-5} 
        DL-MIA & 21.15\% & 11.82\% & 24.02\% & 1.32\% \\
        \bottomrule
    \end{tabular}
    }
    \caption{TPR@1\%FPR for proposed attack and baselines.}
    \label{tab:tpr}
\end{table}

\noindent \textbf{Attack Effectiveness.} Table~\ref{tab:asr} shows ASR values of our proposed MIA and other two baselines. Based on the experimental results, our attack method can distinguish members from non-members with higher ASR value than baselines. Specifically, when the target dataset is MovieLens-1M and DropoutNet is the target recommender system, the ASR value of our proposed attack can be achieved at 0.9340. However, the ASR values of ST-MIA is 0.4995 and ASR value of DL-MIA is 0.5139, which show the performance of baselines are almost equal to random guess. Thus, we claim that existing shadow training based attack methods do not work effectively on hybrid based recommender systems, because the differences between target recommendations and historical interactions for member users and non-member users diminish.

\noindent \textbf{Attack Reliability.} As suggested by~\cite{carlini2022membership}, a powerful and reliable MIA should have a high TPR at a low FPR. Accordingly, we report the TPR values for all methods in our experiments, while fixing the FPR at 1\% following~\cite{fu2024membership}. From the experimental results in Table~\ref{tab:tpr}, we can observe that the TPR values of our proposed method are higher compared with baselines in all experimental settings. For example, when target recommender system is Heater and target dataset is MovieLens-1M, the TPR value of our proposed method achieves at 0.9783. But the performance of baseline attacks is much worse under same experimental settings. Specifically, for ST-MIA, the TPR value is 0.2505, and TPR value is 0.2402 for DL-MIA. The experimental results demonstrate that our proposed attack is more powerful and reliable.

\noindent \textbf{Attack Efficiency.} To compare efficiency of different attack methods, we measure their average computational time across four experimental settings for each method. From Table~\ref{tab:cost time}, our attack takes about 10.4 seconds. In contrast, ST-MIA takes almost 973.3 seconds to implement attack and DL-MIA takes 38,550 seconds to complete attack which is rather expensive in terms of time cost. Instead of directly calculating the value of a metric to infer membership status, these two attacks exploit information from shadow model to train a binary attack model, and adopt a debiasing learning framework in DL-MIA. Experimental results show that our attacks are much more efficient than ST-MIA and DL-MIA which are based on shadow training pipeline.

\begin{table}

    \centering
    
    \resizebox{\linewidth}{!}{
    \begin{tabular}{lccc}
    \toprule
    Method & Our  & ST-MIA & DL-MIA \\
    \midrule
    Time cost (avg) & 10.4 s  & 973.3 s ($\approx$ 93.6 $\times$) & 38,550 s ($\approx$ 3706.7 $\times$) \\
    \bottomrule
     
    \end{tabular}
    }
    
    \caption{The computational cost of our proposed MIAs and baseline attacks, measured by seconds. ($T\times$ represents $T$ times faster of the proposed MIAs than the baselines.)}
    \label{tab:cost time}
\end{table}

\subsection{Why the Proposed MIA Works}

 \begin{figure}
    \centering
    \includegraphics[width=0.8\linewidth]{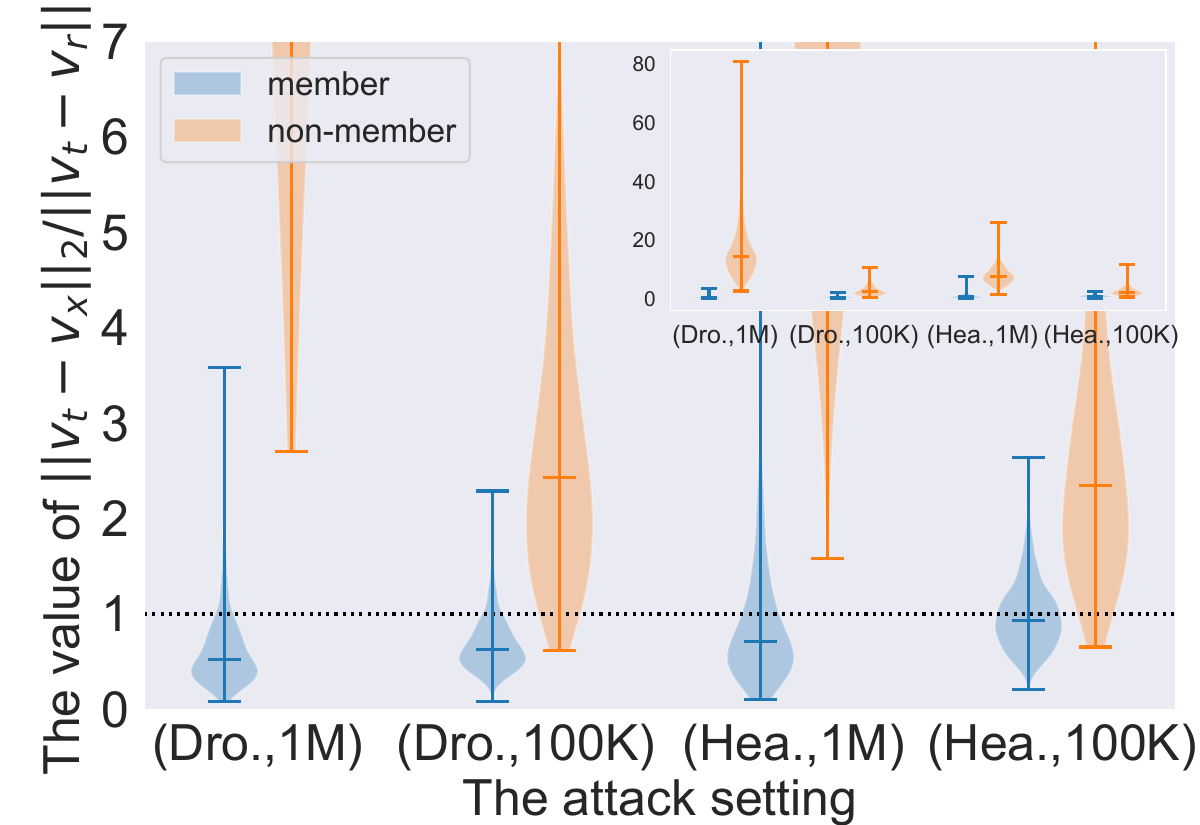}
   \caption{Visualization of $||v_t - v_x||_2 / ||v_t - v_r||_2$ data distribution for members and non-members. }
   \label{fig:violin}
\end{figure}

In our proposed attack, the attacker infers the membership status of a target user's data via comparing the value of $\rho$ with $1$. Thus, to better understand why the proposed attack work, for any target user, we obtain the value of $\rho$, i.e. $\frac{||v_t - v_x||_2}{||v_t - v_r||_2}$, and visualize the distribution of these values for members and non-members in Figure~\ref{fig:violin}. From the visualization, we can observe that the distributions of members and non-members are very different, and the proposed attack achieves a clearer distinction for non-members. Specifically, we can observe that under the attack setting (Dro., 1M), the metric values of all non-members are larger than 1 and values of nearly all members are smaller than 1. In addition, the dotted line almost overlap with the mod-line of data distribution of members in the setting (Hea., 100K). This shows that our proposed attack can predict about 50\% of members correctly in this setting. Furthermore, in the four settings, high portion of non-members can be correctly classified, which shows the reliability of our proposed attack method.

\begin{figure}[t!]
    \centering
  \subfloat[Understanding impact of $n$.]{%
       \includegraphics[width=0.23\textwidth]{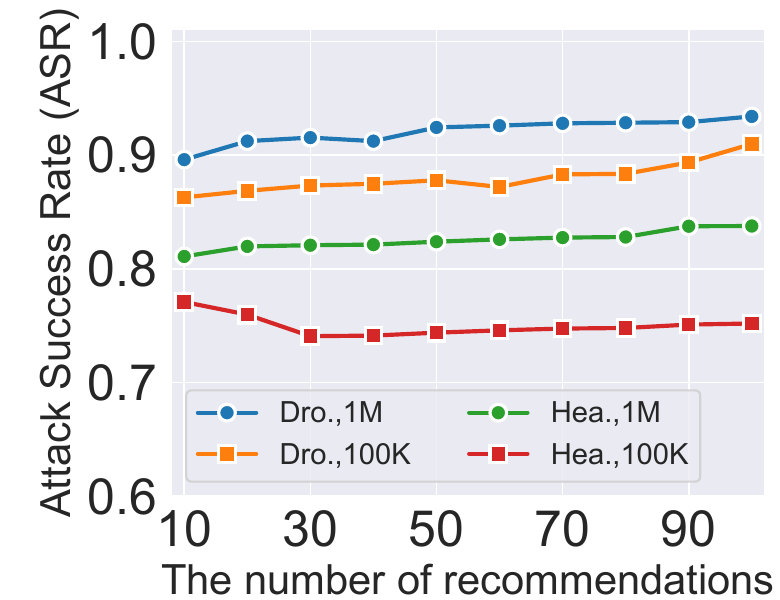}}
  \subfloat[Understanding impact of $l$.]{%
        \includegraphics[width=0.23\textwidth]{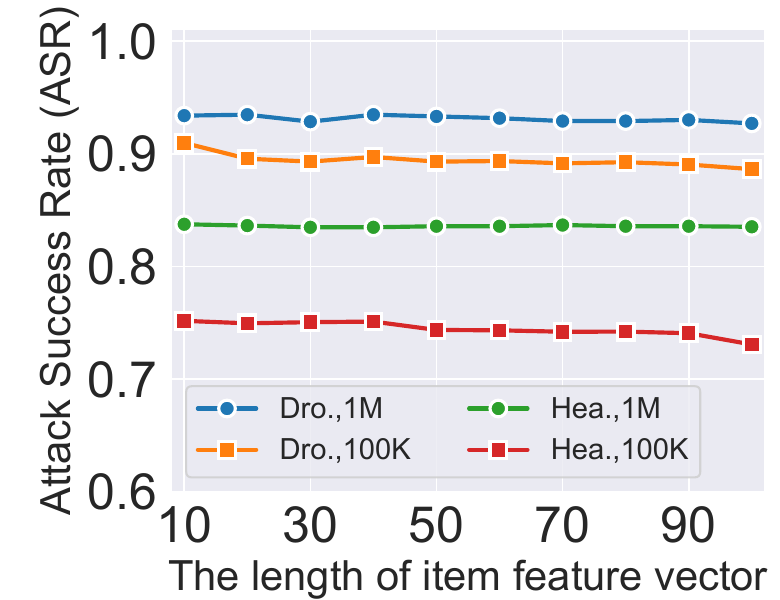}}
  \caption{ASR of proposed attack across different $n$ and $l$.}
  \label{fig:overun}
\end{figure}

 \begin{figure}[t]
    \centering
    \includegraphics[width=0.8\linewidth]{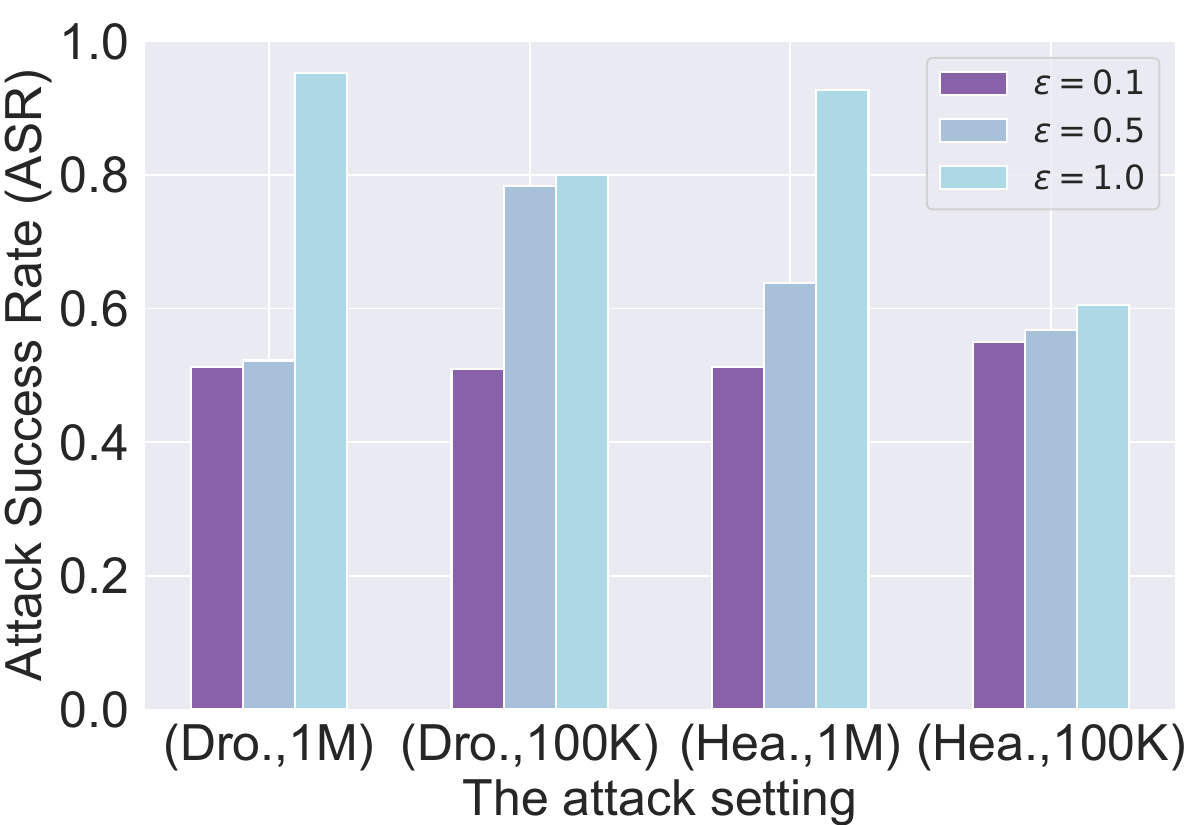}
   \caption{Impact of DP on the ASR of the proposed attack.}
   \label{fig:dp_noise}
\end{figure}

\subsection{Parameter Analysis}

\noindent \textbf{The Number of Recommendations \textit{n}.} We evaluate how the performance of the proposed attack method changes as the number of recommendations $n$ varies. Specifically, we vary the value of $n$ from 10 to 100 to observe the ASR values. Experimental results from Figure~\ref{fig:overun} (a) indicates that the attack performance of our method remains stable as $n$ increases, exhibiting slight improvements as it increases. For instance, the ASR value is less than 0.9 in the setting (Dro., 100K) when $n$ is equal to 10. The ASR value grows stalely and achieves above 0.9 when $n$ is equal to 100. Based on the experimental results presented above, we claim that our proposed attack shows robust performance under the change of the number of recommendations.

\noindent \textbf{The Length of Vectors \textit{l}.} To investigate whether the length of item feature vector $l$ can influence the attack performance, we vary the value of $l$ from 10 to 100. From the experimental results shown in Figure~\ref{fig:overun} (b), we can observe that the parameter has not significant impact on the attack performance. The ASR values show extremely slight decrease tendency with the increase of the length of item feature vector. The ASR values are stable in these attack settings.

\subsection{Defense}

We investigate whether differential privacy (DP)~\cite{dwork2006calibrating} can mitigate the privacy risks of target hybrid-based recommender systems. Existing works~\cite{jagielski2020auditing,choquette2021label} have also applied DP to ML models to mitigate MIAs. Specifically, we focus on user-level DP through adding Gaussian noise on member users' historical interactions to perturb input data of target model, and use ASR to measure the attack performance. In experiments, we follow the parameter settings in~\cite{zhong2024interaction} where $\epsilon$ is relatively set to 0.1, 0.5, 1.0. A smaller $\epsilon$ corresponds to stronger privacy protection. Figure~\ref{fig:dp_noise} reports ASR of our attack for varying $\epsilon$ values. From experimental results, we can observe that ASR values decrease when DP is combined with target model. For example, under (Dro., 100K), ASR reaches 0.5101, 0.7837, and 0.7996 for $\epsilon$ values of 0.1, 0.5, and 1.0, respectively, and rise to 0.9098 without DP. This demonstrates that DP provides a level of privacy protection, while our attack remains consistently effective.

\section{Conclusion}

In this paper, we propose reference recommendation based MIAs along with a relative membership metric to determine the membership status of target users in hybrid-based recommender systems with high efficacy. Unlike existing works assuming unrealistic prior knowledge for the target model to choose different component RSs based on membership, we focus on a more challenging setting where a single hybrid algorithm serves both existing and new users, enabling personalised recommendations based solely on user attributes. A detailed mathematical analysis is provided to highlight the advantages of our proposed metric, with extensive experiments on benchmark datasets and representative hybrid-based recommenders show the effectiveness of our proposed attack. The findings in this paper offer new insights into the privacy vulnerabilities of modern recommender systems.

\clearpage

\section{Acknowledgments}
The work was partially supported by The State Key Laboratory of Novel Software Technology (KFKT2024A03). This work was also partially supported by the Australian Research Council Discovery Project DP230100676.

\bigskip

\bibliography{aaai2026}

\end{document}